\documentstyle[epsf,graphics,epsfig,times]{mn}

\newcommand{\msun}{\thinspace\hbox{${\rm M}_{\odot}$}\ }

\def\simgt{\mathrel{\lower0.6ex\hbox{$\buildrel {\textstyle >}
 \over {\scriptstyle \sim}$}}}
\def\simlt{\mathrel{\lower0.6ex\hbox{$\buildrel {\textstyle <}
 \over {\scriptstyle \sim}$}}}

\title[The clustering of halo mergers] 
{The clustering of halo mergers}

\author[Will J.\ Percival et al.]{
\parbox[t]{\textwidth}{
Will J.\ Percival$^1$,
Douglas Scott$^2$,
John A.\ Peacock$^1$,
James S.\ Dunlop$^1$}
\vspace*{6pt} \\
$^1$Institute for Astronomy, University of Edinburgh, Royal Observatory, 
       Blackford Hill, Edinburgh EH9 3HJ, UK \\
$^2$Department of Physics and Astronomy, University of British
       Columbia, Vancouver, BC, Canada V6T 1Z1 \\
}
\date{Submitted for publication in MNRAS}
\begin{document}
\maketitle

\begin{abstract}
We analyse the spatial distribution of halo merger sites using four
numerical simulations with high time resolution of structure growth in
a $\Lambda$-CDM cosmology. We find no evidence for any large-scale
relative bias between mergers and randomly selected haloes of the same
mass at high redshift. Given a sample of galaxies that form a Poisson
sampling of halo mergers, then the amplitude of the measured
clustering ought to lead to a robust estimate of the median host halo
mass. In the hierarchical picture of structure growth, mergers between
galaxies at high redshifts are thought to create dust enshrouded
starbursts leading to increased emission in the sub-mm
wavebands. Hence, a measurement of the clustering strength of SCUBA
galaxies, for example, can determine whether or not they are giant
elliptical galaxies in the process of formation.
\end{abstract}


\section{Introduction}  \label{sec:intro}

\begin{figure}
  \setlength{\epsfxsize}{\columnwidth} \centerline{\epsfbox{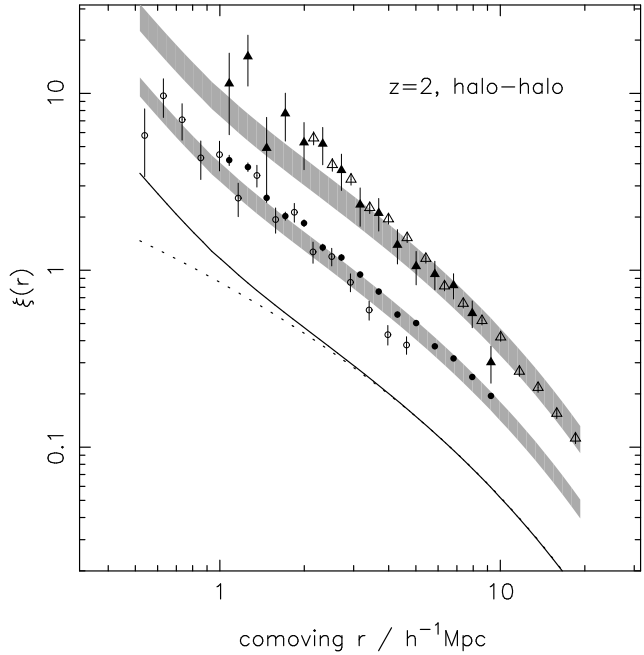}}

  \caption{ The halo-halo correlation function as determined from a
  redshift slice at $z=2.0$, calculated for all haloes with mass
  (2.9--5.7)$\times10^{11}\msun$ and (2.3--4.5)$\times10^{12}\msun$
  found in the $\Lambda$CDM$_{100a}$ simulation, corresponding to
  groups with 41--80 particles (solid triangles) or 321--640 particles
  (solid circles). For comparison we also plot the distribution of
  haloes in the lower mass range found in the $\Lambda$CDM$_{50}$
  simulation (321--640 particles, open circles), and similarly the
  distribution of haloes in the upper mass range found in the
  $\Lambda$CDM$_{200}$ simulation (41--80 particles, open
  triangles). The correlation functions were determined in bins of
  equal width in $\ln r$ from 1 to 10 per cent of the box width. The
  dotted line shows the expected linear correlation function of the
  mass, while the solid line shows that determined using the
  non-linear power spectrum fitting formulae of Smith et
  al.~(2002). The grey shaded regions show the expected halo-halo
  correlation functions for the two mass ranges calculated using the
  fit of Smith et al.~(2002) together with the bias formula of Sheth
  \& Tormen (1999). As can be seen, the peak-background split works
  remarkably well in the linear regime, while there is weak evidence
  that either the bias becomes scale dependent in the quasi-linear
  regime, or the wavelength of quasi-linear behaviour changes with
  halo properties. \label{fig:xi_mass}}
\end{figure}

\begin{figure}
  \setlength{\epsfxsize}{\columnwidth} \centerline{\epsfbox{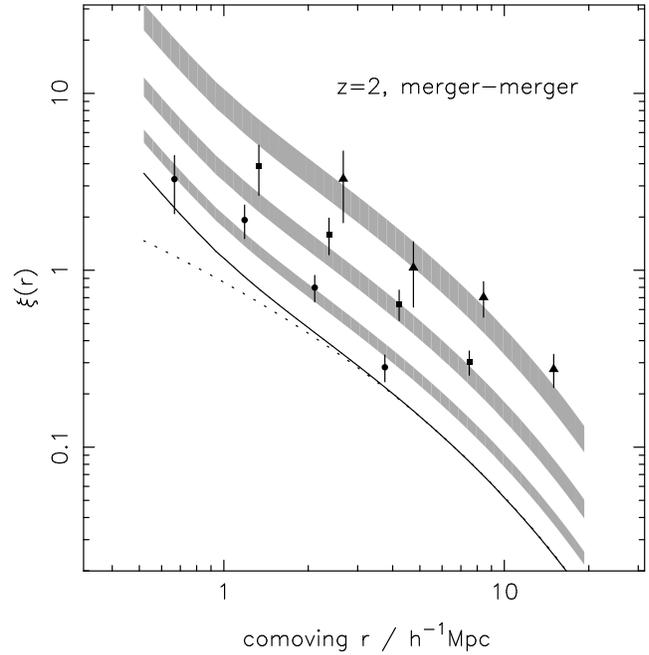}}

  \caption{ The merger-merger correlation function for haloes in merger
  Subset~2 (see Section~\ref{sec:sim} for details) as determined at
  $z=2.0$ for haloes with (3.6--7.1)$\times10^{10}\msun$ (solid circles),
  (2.9--5.7)$\times10^{11}\msun$ (solid
  squares) and (2.3--4.5)$\times10^{12}\msun$
  (solid triangles), found in the $\Lambda$CDM$_{50}$,
  $\Lambda$CDM$_{100a}$ \& $\Lambda$CDM$_{200}$ simulations
  respectively. These mass ranges correspond to groups with $41-80$
  particles in each simulation. The solid and dotted lines are as in
  Fig.~\ref{fig:xi_mass}, and the grey shaded regions show the
  predicted correlation functions using the bias model of Sheth \&
  Tormen (1999) for the mass ranges plotted, again as in
  Fig.~\ref{fig:xi_mass}. For comparison with Fig.~\ref{fig:bias_all},
  in this plot $1.23<\nu<2.42$. We see no significant difference
  between the correlation functions for the subset of haloes that have
  undergone recent mergers and that of all haloes. \label{fig:xi_merg}}
\end{figure}

A great deal of observational effort has been expended in measuring
the clustering of various subsets of high-redshift galaxies. Of these,
the most successful have been surveys conducted for QSOs (e.g.~Croom
et al.~2001), Lyman-break galaxies at $z\simeq3$ (Giavalisco et
al.~1998; Porciani \& Giavalisco 2002), and $z\simeq4$ (Ouchi et
al.~2001), and EROs (Daddi et al.~2000). A particularly exciting new
area is the analysis of sub-mm galaxies, made possible by the SCUBA
instrument on the James Clerk Maxwell Telescope (Holland et al.~1999).
Some of the most recent surveys have provided tantalising evidence
that luminous sub-mm galaxies are also strongly clustered (Scott et
al. in prep.; Webb et al.~2002).  Planned new surveys, particularly
the recently commenced SHADES (SCUBA HAlf Degree Extragalactic Survey,
{\tt http://www.roe.ac.uk/ifa/shades}) project will extend this
analysis by finding $200-400$ luminous sub-mm sources in a half-degree
field. The SHADES data will cover separations of order
0.1--$10\,h^{-1}{\rm Mpc}$ at redshifts 2--4, so the clustering of
objects found at these redshifts will be measured around the
transition from the linear to quasi-linear regimes (see
Figs.~\ref{fig:xi_mass} and \ref{fig:xi_merg}). Following the halo
model (Jing, Mo \& Boerner 1998; Seljak 2000; Peacock \& Smith 2000;
Ma \& Fry 2000), this large-scale clustering is expected to be
dominated by halo-halo clustering and the number of galaxies within
each halo simply weights the relative importance of haloes with a
given mass.  A large-scale clustering study also avoids the additional
complications involved in relating the halo properties to luminosity,
since the details of the generation of radiation will mainly affect
the counts and background values, rather than the fluctuations (see
e.g.~Knox et al.~2001). The measured clustering is therefore only
dependent on the distribution of host halo masses and on any intrinsic
properties of the haloes caused by their evolutionary history.

The small overlap between sub-mm and X-ray selected samples (Barger et
al.~2001; Almaini et al.~2002) discourages models of sub-mm galaxies
being buried AGN and favours dust enshrouded starbursts.  Mergers
between galaxies are often invoked as the driving mechanism for the
luminous emission as they are known to cause such star-bursting
activity. Detailed simulations of isolated halo mergers suggest that
bursts of star formation associated with mergers between either disk
or bulge/halo systems with a variety of relative masses all have
lifetimes which are ${<}\,10^8$\,years (e.g.~Mihos \& Hernquist
1996). Additionally, such mergers predominantly occur at high redshift
(Lacey \& Cole 1993; Percival, Miller \& Peacock 2000), fitting in
with our present knowledge of SCUBA sources (Blain et al.~2002). One
of the simplest viable models for the distribution of sub-mm sources
is therefore that they comprise a Poisson sampling of the distribution
of halo-halo mergers. This ties in with proposed models of Lyman-break
galaxies at $z\sim3$, in which the emission is driven by either
merger-induced starbursts (Kolatt et al. 1999) or quiescent star
formation (Baugh et al. 1998). A recent comparison of the clustering
predicted by various models was given by Wechsler et al. (2001).

This letter focuses on halo properties and considers the importance of
halo mergers for clustering strength. The clustering of dark-matter
haloes in numerical simulations has been studied in detail
(e.g. Colberg et al. 2000, who analysed the Hubble volume simulation
at $z=0$). There has also been previous work looking at the clustering
of haloes as a function of their intrinsic properties. Knebe \& Muller
(1999) showed that non-virialised haloes (plausibly those with an
ongoing merger) found using a friends-of-friends (FOF) algorithm are
more biased than virialised ones. Recent work by Gottl{\"o}ber et
al.~(2002) also seems to find that mergers can alter the bias of
haloes. However, in Gottl{\"o}ber et al.~(2002), and in the
Lyman-break galaxy model comparison of Wechsler et al. (2001), the
distribution of host halo masses was not fixed between models. It is
therefore not possible to distinguish the relative importance of halo
mass and intrinsic halo properties to the clustering strength. Using
numerical simulations, Lemson \& Kauffmann (1999) found that only the
mass function varied as a function of environment, while the formation
redshift, concentration, shape and spin of each halo were independent
of the environment. It is plausible that recent mergers would have
affected the halo properties that they considered, and so this work
implies that halo mergers should cluster in the same way as haloes in
general. 

In a recent paper, Kauffmann \& Haehnelt (2002) analysed the
distribution of quasars in a model that combined a cosmological N-body
simulation with a simple merger-based prescription for AGN
activation. Their paper considered the relative bias of quasars and
galaxies over a range of scales, redshifts and quasar and galaxy
luminosities. As part of their analysis, Kauffmann \& Haehnelt also
presented some evidence that the large-scale clustering of haloes with
recent mergers followed that of the mean halo population of the same
mass on large scales.

Given the potential importance of halo mergers for both quasar
activation as discussed by Kauffmann \& Haehnelt (2002) and
merger-induced starbursts as discussed above, this letter extends
previous work by using four numerical simulations with high time
resolution of regions within a $\Lambda$-CDM cosmology to provide a
detailed analysis of the large-scale clustering of mergers. The layout
is as follows. In Section~\ref{sec:sim} we describe the parameters of
the simulations used and the method adopted to calculate the bias of
each subset of haloes, while in Section~\ref{sec:models} we briefly
introduce analytic halo bias models. Results are are presented in
Section~\ref{sec:res}, and we end with a discussion of these results
in Section~\ref{sec:dis}.

\section{The simulations} \label{sec:sim}

We have completed four $N$-body simulations with different mass
resolution using {\sc gadget}, a parallel tree code (Springel, Yoshida
\& White 2001). Each simulation contained $256^3$ particles initially
distributed using the {\sc cosmics} package ({\tt
http://arcturus.mit.edu/cosmics/}), altered to use the transfer
function fitting formulae of Eisenstein \& Hu (1999). Cosmological
parameters were fixed at $\Omega_{\rm m}=0.3$, $\Omega_{\rm
b}/\Omega_{\rm m}=0.15$, $\Omega_\Lambda=0.7$, $h=0.7$, $n_{\rm
s}=1$. Other simulation parameters are given in
Table~\ref{tab:sim}. Note that although the power spectrum allowed for
the signature of baryons, the particles in the simulation were all
collisionless. These four complementary simulations cover a range in
mass at $\sigma_8=0.75$, and cover two different power spectrum
normalisations at fixed particle mass.

\begin{table}
  \centering
  \caption{Normalization, $\sigma_8$, box size, particle mass $M_{\rm
  part}$, gravitational softening radius $r_{\rm soft}$, and number of
  outputs $N_{\rm out}$, for the four N-body simulations considered in
  this work. \label{tab:sim}}

  \begin{tabular}{@{}llcccc@{}}  \hline 
  simulation & $\sigma_8$ & box size & $M_{\rm part}$ & $r_{\rm soft}$
    & $N_{\rm out}$ \\
       & & $h^{-1}{\rm Mpc}$ & \msun & $h^{-1}{\rm kpc}$ & \\ \hline 
  $\Lambda$CDM$_{050}$  & 0.75 & 50  & $8.87\times10^{8}$  & 10 & 453 \\
  $\Lambda$CDM$_{100a}$ & 0.75 & 100 & $7.10\times10^{9}$  & 20 & 426 \\
  $\Lambda$CDM$_{200}$  & 0.75 & 200 & $5.68\times10^{10}$ & 40 & 402 \\
  $\Lambda$CDM$_{100b}$ & 0.9  & 100 & $7.10\times10^{9}$  & 20 & 434 \\ \hline 
  \end{tabular}
\end{table}

Halo groups were found at $400-450$ epochs in each simulation,
separated by approximately equal intervals in $\ln a$, using a
standard friends-of-friends algorithm with $b=0.2$. No attempt was
made to add dynamical criteria, such as removing unbound
particles. Because of the group finding algorithm used, it is
important to realise that the haloes considered in this work are
isolated and that the halo mass is a measure of the total mass of the
system. For each halo we determine the centre of mass and use this in
subsequent calculations of clustering properties. We used both the
power spectrum and the correlation function to quantify clustering and
have also considered a number of different estimators to calculate the
correlation function with or without using the periodic nature of the
simulations. All of these methods produced consistent results.

In the following we consider four samples of haloes: the set of {\em
all\/} haloes, and three subsamples with special properties.  The
first subset consists of all {\em new\/} haloes, defined to be haloes
with more than 50 per cent of constituent particles that had not been
found in a halo of equal or greater mass at an earlier epoch. This is
the set of haloes that have grown by at least a single particle
between simulation output, and is designed to limit numerical effects:
each halo is only considered to have reached a mass $M$ at a single
epoch. In the following, halo 1 (mass $M_1$ at time $t_1$) is
considered to be a progenitor of new halo 2 (mass $M_2>M_1$ at time
$t_2>t_1$) if at least 50 per cent of the particles in halo 1 are
contained in halo 2.

To distinguish between mergers and haloes formed by slow accretion, we
consider two subsamples of new haloes that have undergone a recent
merger. In order to obtain sufficient mergers to measure the
clustering, we consider all mergers that happened within $10^8$\,years
of a given epoch. The exact form of this lifetime is not significant
for the results presented in this letter and we note that any
observational signature resulting from a merger would be present for
such a finite timescale. Subset~2 consists of all haloes that were
defined as new within $10^8$\,yrs and had two progenitors at the
previous time step with masses between 25 per cent and 75 per cent of
the final mass.  Hence this consists of haloes that have undergone a
recent violent merger, in that two progenitors with a 3:1 mass ratio
or less, making up $>50$ per cent of the total mass of the halo,
merged within the last $10^8$\,yrs. The last subsample (Subset~3) is
similar except that we allow all new haloes with two progenitors with
masses between 15 per cent and 85 per cent of the final mass (i.e. a
merger with a mass ratio of 17:3 or less occurred within
$10^8$\,yrs). Due to the discrete nature of our knowledge of halo
growth in time, this procedure may miss halo mergers where a
progenitor forms and merges between outputs from the
simulation. However, we do not expect any time-selection effects
induced by this sampling to bias the spatial distribution of our
sample.  Before discussing the results for these four halo samples, we
first discuss a simple model for predicting halo biasing, which we
will use for comparison.

\section{Analytic models of halo bias}  \label{sec:models}

In this Section we review the peak-background split model for the bias
of the mean halo population, a model that we will use later to compare
with the large-scale bias of halo mergers recovered from numerical
simulations. In the peak-background split model (Cole \& Kaiser 1989;
Mo \& White 1996; Sheth \& Tormen 1999), the density field is divided
into two components: $\delta$ which contains the small-scale
fluctuations from which a given halo forms; and an underlying
large-scale density field $\epsilon$. The effect of the large-scale
density field is to perturb the number density $n(M)$ of haloes
leading to a Lagrangian bias $\Delta n(M)/n(M)=b_{\rm L}\epsilon$. In
addition, the growth of the background field $\epsilon$ leads to an
increase in the clustering: for small $\epsilon$ this gives an
Eulerian bias $b_{\rm E}=1+b_{\rm L}$. The result of this model is
that the predicted bias for small $\epsilon$ is dependent on the
derivative of the logarithm of the halo mass function with respect to
the density. In the following we shall only consider the simple model
of linear biasing that follows from this analysis, assuming that
$\xi_{\rm hh}=b_{\rm E}^2\xi_{\rm mm}$ and $P_{\rm hh}(k)=b_{\rm
E}^2P_{\rm mm}(k)$ in the linear regime, where $\xi_{\rm hh}$ and
$\xi_{\rm mm}$ are the halo and mass correlation functions and $P_{\rm
hh}(k)$ and $P_{\rm mm}(k)$ are the halo and mass power spectra.

Following standard convention, we define $\nu\equiv\delta_{\rm
c}/\sigma_M$ as the ratio of the density for collapse, calculated
using the top-hat collapse model, to the expected rms fluctuations in
top-hat spheres containing average mass $M$. There is strong evidence
that the mass function determined from numerical simulations is a
function of $\nu$ alone for large $\nu$ (e.g.~Jenkins et al.~2001),
which implies that the bias should also be a function of $\nu$
only. In the following we consider the bias calculated from the
standard Press--Schechter mass function (Press \& Schechter 1974; Mo
\& White 1996), and from fits to the mass function recovered from
numerical simulations by Sheth \& Tormen (1999), and by Jenkins et
al. (2001). The fit to the mass function presented by Sheth \& Tormen
(1999) has been shown to be close to that recovered using non-standard
versions of Press--Schechter theory that involve either replacing the
spherical top-hat collapse model with an ellipsoidal collapse model
(Sheth, Mo \& Tormen 2001), or calculating the mass function using the
spherical top-hat filter rather than the sharp $k$-space filter
(Percival 2001).

\begin{figure*}
  \setlength{\epsfxsize}{1.0\textwidth}
  \centerline{\epsfbox{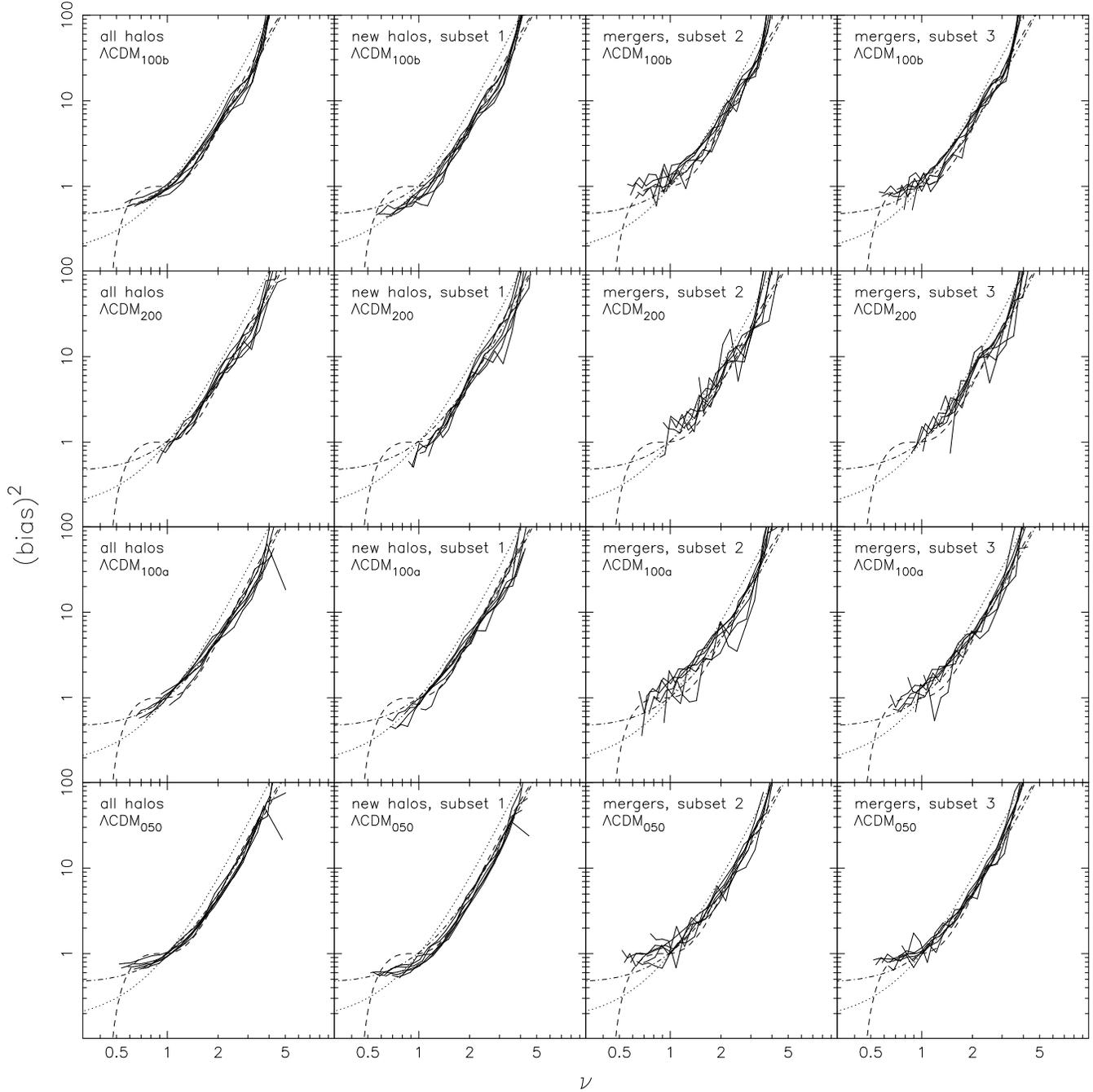}} \caption{The Eulerian bias
  squared as a function of $\nu\equiv\delta_c/\sigma_M$, calculated
  for 6 mass ranges corresponding to groups of $21$--$30$, $31$--$40$,
  $41$--$80$, $81$--$160$, $161$--$320$ and $321$--$640$ particles
  (solid lines) in each of the four simulations (rows of panels).
  Each line corresponds to the evolution of the clustering of haloes
  with time, and was calculated by comparing the power spectra of the
  set of haloes and of the mass. Power spectra were calculated from a
  $128^3$ Fourier transform covering each simulation cube, and the
  power was compared at wavelengths less than $0.03$ times the Nyquist
  frequency. This was chosen as a compromise between obtaining
  sufficient signal and avoiding the onset of quasi-linear
  behaviour. The power spectra calculated for each set of haloes were
  averaged into 21 bins in $\ln a$ before comparison with the mass
  spectra. The data are compared to analytic models of the bias
  resulting from the standard Press--Schechter mass function (dotted
  line), the Sheth \& Tormen (1999) mass function (dot-dash line) and
  the Jenkins et al. (2001) mass function (dashed line). Although the
  Jenkins et al.~(2001) mass function was calculated for $\nu>0.5$,
  they only had significant data at higher $\nu$, so the strong
  decrease of the bias predicted using their mass function at small
  $\nu$ is probably not important.  \label{fig:bias_all}}
\end{figure*}

\section{Results}  \label{sec:res}

Fig.~\ref{fig:xi_mass} shows the halo-halo correlation functions for
two mass ranges calculated from the redshift slice at $z=2$. These
functions are compared to, and shown to be in good agreement with the
analytic model of bias calculated from the Sheth \& Tormen (1999) mass
function as discussed in Section~\ref{sec:models}, and consistent
results are shown to be recovered from the three simulations with
$\sigma_8=0.75$ for the same mass ranges. Fig.~\ref{fig:xi_merg} shows
an example correlation function, again calculated at $z=2$, but now
only considering haloes that have undergone a recent merger -- Subset~2
(described in Section~\ref{sec:sim}). Here, we have considered three
mass ranges corresponding to groups of $41-80$ particles in each of
the $\sigma_8=0.75$ simulations.
We find that this subset of haloes is also distributed
in a consistent way to that expected from the analytic models.

In Fig.~\ref{fig:bias_all} we plot the linear Eulerian bias squared as
a function of $\nu$ for the four subsamples of haloes described in
Section~\ref{sec:sim}.  We consider six mass ranges, corresponding to
groups of $21$--$30$, $31$--$40$, $41$--$80$, $81$--$160$,
$161$--$320$ and $321$--$640$ particles in each of the four
simulations. Each solid line, of which there are six in each panel,
corresponds to one of these mass ranges, and for such lines, $\nu$ can
be considered a function of time only. The wavelength at which
quasi-linear behaviour becomes dominant in the halo-halo power
spectrum is expected to be a function of both the epoch and halo mass
selected. For example, the first haloes to form at high redshift will
be in high density regions that are more likely (than a mass element
selected at random) to have broken free from linear growth on a larger
scale. Given no available model for the linear cut-off wavelength for
each set of haloes, the bias was calculated from the ratio of halo and
mass power spectra for $k$ less than a fixed limit. As a compromise
between staying below the quasi-linear cutoff and obtaining sufficient
signal in the calculated power spectra, the limit used was set at
$0.03$ times the Nyquist frequency of the $128^3$ Fourier transforms
performed for each simulation box. For each panel, the evolution of
the bias as a function of time can be seen to be well described by the
peak-background split model. The slight turn up at large $\nu$ away
from the models is consistent with the quasi-linear epoch entering the
range of $k$-values considered -- increasing this limit decreases the
$\nu$ at which this effect starts.

\section{Discussion}  \label{sec:dis}

Fig.~\ref{fig:bias_all} shows that the linear bias squared calculated
from all four sets of haloes and all simulations follows the
predictions of the peak-background split calculated using the Sheth \&
Tormen (1999) fit to the simulation mass function remarkably
well. Comparing the bias recovered from the set of all haloes with
previous results, we find both the decrease away from the
Press--Schechter prediction at $\nu>1$ and a turn-up in the bias at
$\nu<1$ found by Jing (1998; 1999) and by Sheth \& Tormen
(1999). These features are replicated in the three subsamples of
haloes that we have considered, although there is weak evidence that,
for $\nu<1$, the set of new haloes is less biased, and the sets of
mergers more biased than the set of all haloes. However, these
deviations are modest, and do not affect the primary conclusions of
this letter.

Our motivation has been to provide a first step for models of the
clustering of relatively small samples of high redshift objects.  At
such redshifts we find no evidence for a significant difference
between the clustering of the set of all haloes of a given mass and
any of the three subsets that we have chosen. A systematic deviation
could of course be hidden within the noise, but would have to be at a
level well below our current knowledge of the clustering properties of
high redshift galaxies ($\simlt$\,20 per cent in the bias). A
consistent environment for mergers, compared with the mean halo
population, is in agreement with the work of Lemson \& Kauffmann
(1999) and with Press--Schechter theory with a sharp $k$-space filter,
which predicts that the build-up of a halo around a small mass element
is Markovian in nature. The environment of a halo at time $t$ and mass
$M$ is determined by the behaviour of the trajectory at mass $>M$
which is independent of the behaviour $<M$ that contains details of
the halo properties.

Several different classes of galaxy, detected at a variety of
wavebands, could be related to merging haloes.  Determining how the
mergers relate to such luminous objects requires a great deal of
additional modelling.  Work along these line is still quite
phenomenological in nature, although some progress has been made
(e.g.~Kauffmann \& Haehnelt~2000; Silva et al.~2001; van Kampen et
al.~in preparation). However, it appears from our study that to a
reasonable degree of approximation the amplitude of clustering will be
a direct measure of the mass of the underlying haloes. So if we focus
on SCUBA galaxies for example, a reliable measurement of the
clustering amplitude should make it clear whether the objects involved
are giant elliptical galaxies in the process of formation or related
to much less massive objects.

\section{Acknowledgements}
We are indebted to Volker Springel for making the {\sc gadget} N-body
code (Springel, Yoshida \& White 2001) publicly available. DS is
supported by the Natural Sciences and Engineering Research Council of
Canada, and thanks the IfA for their hospitality while this work was
carried out. JSD \& JAP are grateful for the support of PPARC Senior
Research Fellowships.


\begin{thebibliography}{}
  \bibitem[\protect\citename{}]{Almetal}
    Almaini O., et al., 2001, MNRAS, submitted [astro-ph/0108400]
  \bibitem[\protect\citename{}]{Baretal}
    Barger A.J., Cowie L.L., Steffen A.T., Hornschemeier A.E., Brandt W.N.,
     Garmire G.P., 2001, ApJ, 560, L23
  \bibitem[\protect\citename{}]{baugh}
    Baugh C.M., Cole S., Frenk C.S., Lacey C.G., 1998, ApJ, 498, 504
  \bibitem[\protect\citename{}]{Blaetal}
    Blain A.W., Smail I., Ivison R.J., Kneib J.-P., Frayer D.T., 2002,
     Phys. Rep., in press [astro-ph/0202228]
  \bibitem[\protect\citename{}]{colberg} 
    Colberg J.M., et al. (The Virgo Consortium), 2000, MNRAS, 319, 209
  \bibitem[\protect\citename{}]{cole} 
    Cole S., Kaiser N., 1989, MNRAS, 237, 1127
  \bibitem[\protect\citename{}]{Croetal}
    Croom S.M., Shanks T., Boyle B.J., Smith R.J., Miller L., Loaring N.S.,
    Hoyle F., 2001, MNRAS, 325, 483
  \bibitem[\protect\citename{}]{daddi} 
    Daddi E., et al., 2000, A\&A, 361, 535
  \bibitem[\protect\citename{}]{Eisenstein} 
    Eisenstein D.J., Hu W., 1998, ApJ, 496, 605
  \bibitem[\protect\citename{}]{Giaetal}
    Giavalisco M., Steidel C.C., Adelberger K.L., Dickinson M.E., Pettini M.,
     Kellogg M., 1998, ApJ, 503, 543
  \bibitem[\protect\citename{}]{Holetal}
    Holland W.S., et al., 1999, MNRAS, 303, 659
  \bibitem[\protect\citename{}]{Gottloeber}
    Gottloeber S., Kerscher M., Kravtsov A.V., Faltenbacher A., Klypin
    A., Muller V., 2002, A\&A, 387, 778
  \bibitem[\protect\citename{}]{jenkins}
    Jenkins A., Frenk C.S., White S.D.M., Colberg J.M., Cole S.,
    Evrard A.E., Couchman H.M.P., Yoshida N., 2001, MNRAS, 321, 372
  \bibitem[\protect\citename{}]{jing3}
    Jing Y.P., Mo H.J., Boerner G., 1998, ApJ, 494, 1
  \bibitem[\protect\citename{}]{jing1}
    Jing Y.P., 1998, ApJ, 503, L9
  \bibitem[\protect\citename{}]{jing2}
    Jing Y.P., 1999, ApJ, 515, L45
  \bibitem[\protect\citename{}]{KauHae1}
    Kauffmann G., Haehnelt M., 2000, MNRAS, 311, 576
  \bibitem[\protect\citename{}]{KauHae2}
    Kauffmann G., Haehnelt M., 2002, MNRAS, 332, 529
  \bibitem[\protect\citename{}]{Knoetal}
    Knox L., Cooray A., Eisenstein D., Haiman Z., 2001, ApJ, 550, 7
  \bibitem[\protect\citename{}]{Knebe}
    Knebe A., Muller V., 1999, A\&A, 341, 1
  \bibitem[\protect\citename{}]{Kolatt}
    Kolatt et al., 1999, ApJ, 523, L109
  \bibitem[\protect\citename{}]{lc93} 
    Lacey C., Cole S., 1993, MNRAS, 262, 627  
  \bibitem[\protect\citename{}]{lk} 
    Lemson G., Kauffmann G., 1999, MNRAS, 302, 111
  \bibitem[\protect\citename{}]{ma} 
    Ma C.-P., Fry J.N., 2000, ApJ, 543, 503
  \bibitem[\protect\citename{}]{MihHer96}
    Mihos J.C., Hernquist L., 1996, ApJ, 464, 641
  \bibitem[\protect\citename{}]{Mo}
    Mo H.J., White S.D.M., 1996, MNRAS, 282, 347
  \bibitem[\protect\citename{}]{Oucetal}
    Ouchi M., et al., 2001, ApJ, 558, L83
  \bibitem[\protect\citename{}]{peacock} 
    Peacock J.A., Smith R.E., 2000, MNRAS, 318, 1144
  \bibitem[\protect\citename{}]{percival1} 
    Percival W.J., Miller L., Peacock J.A., 2000, MNRAS 318, 273
  \bibitem[\protect\citename{}]{percival2} 
    Percival W.J., 2001, MNRAS, 327, 1313
  \bibitem[\protect\citename{}]{PorGia}
    Porciani C., Giavalisco M., 2002, ApJ, 565, 24
  \bibitem[\protect\citename{}]{ps} 
    Press W., Schechter P., 1974, ApJ, 187, 425 
 \bibitem[\protect\citename{}]{seljak} 
    Seljak U., 2000, MNRAS, 318, 203
 \bibitem[\protect\citename{}]{sheth99} 
    Sheth R.K., Tormen G., 1999, MNRAS, 308, 119
 \bibitem[\protect\citename{}]{sheth01} 
    Sheth R.K., Mo H.J., Tormen G., 2001, MNRAS, 323, 1
  \bibitem[\protect\citename{}]{smith} 
    Smith R.E., et al. (The Virgo Consortium), 2002, MNRAS submitted
     [astro-ph/0207664]
  \bibitem[\protect\citename{}]{Siletal}
    Silva L., Granato G.L., Bressan A., Lacey C., Baugh C.M., Cole S.,
     Frenk C.S., 2001, Ap\&SS, 276, 1073
  \bibitem[\protect\citename{}]{Springel}
    Springel V., Yoshida N., White S.D.M., 2001, NewA, 6, 79
  \bibitem[\protect\citename{}]{Webetal}
    Webb T.M.A., Eales S.A., Lilly S.J., Clements D.L., Dunne L., Gear W.K.,
     Flores H., Yun M., 2002, ApJ, submitted [astro-ph/0201180]
  \bibitem[\protect\citename{}]{weschler}
    Weschler R.S., Somerville R.S., Bullock J.S., Kolatt T.S., Primack
    J.R., Blumenthal G.R., Dekel A., 2001, ApJ, 554, 85 
\end{thebibliography}
\end{document}